\documentclass{aa}
\usepackage[varg]{txfonts}
\usepackage{graphicx}
\usepackage{t1enc}
\usepackage{epsfig}
\usepackage{rotating}
\usepackage{natbib}
\bibpunct{(}{)}{;}{a}{}{,}

\newcommand{\Tkin}	{$T_\mathrm{kin}$}
\newcommand{\mum}   {$\mu$m}
\newcommand{\kms}{\hbox{\kern 0.20em km\kern 0.20em s$^{-1}$}}
\newcommand{\cmt}{\hbox{\kern 0.20em cm$^{-3}$}}
\newcommand{\cmd}{\hbox{\kern 0.20em cm$^{-2}$}}

\newcommand{\lo}    {$L_{\sun}$}

\newcommand{\nh}    {NH$_3$}

\newcommand{\chtoh} {CH$_3$OH}
\newcommand{\hho}   {H$_2$O}

\newcommand{\hh}   {H$_2$}

\newcommand{\et}    {et al.}
\newcommand{\eg}    {e.\,g.,}
\newcommand{\ie}     {i.\,e.,}

\newcommand{\vel} {$v_\mathrm{LSR}$}
\newcommand{\velo} {$v$}

\newcommand{\hcop}  {HCO$^+$}

\newcommand{\tco}   {$^{13}$CO}

\newcommand{\phn}   {\phantom{0}}
\newcommand{\phnn}  {\phantom{0}\phantom{0}}

\newcommand{\phs}   {\phantom{$^0$}}

\newcommand{\phe}   {\phantom{$^\mathrm{c}$}}

\newcommand{\phwater} {\phantom{$o$-H$_2$O}}

\newcommand{\chtcn}   {CH$_3$CN}

\begin{document}

	\title{The CHESS survey of the L1157-B1 bow-shock: high and low excitation water vapor \thanks{Based on \textit{Herschel} HIFI and PACS observations. \textit{Herschel}
is an ESA space observatory with science instruments provided by European-led Principal Investigator consortia and with important participation
from NASA.}}

	\author{G. Busquet\inst{1}
       	 \and
	  B. Lefloch\inst{2,3}
        \and
	M. Benedettini \inst{1}
	 \and
	C. Ceccarelli\inst{2}
	\and
	C. Codella\inst{4}
	\and
        S. Cabrit	\inst{5}
        \and
                B. Nisini		\inst{6}
                \and
                S. Viti	\inst{7}
                \and
                  	A.~I. G\'omez-Ruiz	\inst{4}
	\and
	A. Gusdorf \inst{8}
	\and
	A.~M. di Giorgio	\inst{1}
	\and
	 L. Wiesenfeld	\inst{2}
         }

  	\offprints{Gemma Busquet,\\ \email{gemma.busquet@iaps.inaf.it}}

	\institute{INAF - Istituto di Astrofisica e Planetologia Spaziali. Via del Fosso del Cavaliere 100, I-00133 Roma, Italy
        \and
	 UJF-Grenoble 1 / CNRS-INSU, Institut de Plan\'etologie et d'Astrophysique de Grenoble (IPAG) UMR 5274, Grenoble F-38041, France
	 \and
	 Centro de Astrobiologia, CSIC-INTA, Carretera de Torrej\'on a  Ajalvir, km 4, Torrej\'on de Ardoz, E-28850 Madrid, Spain
	  \and
	 INAF, Osservatorio Astrofisico di Arcetri, Largo Enrico Fermi 5, I-50125 Firenze, Italy
	  \and
        Observatoire de Paris, LERMA,  UMR 8112 du CNRS, ENS, UPMC, UCP,  61 Av. de l'Observatoire, F-75014 Paris, France
	 \and
	INAF-Osservatorio Astronomico di Roma, Via di Frascati 33, I-00040 Monte Porzio Catone, Italy
	\and
	Department of Physics and Astronomy, University College London, Gower Street, London, WC1E 6BT, UK
	\and
	LERMA, UMR 8112 du CNRS, Observatoire de Paris, \'Ecole Normale Sup\'erieure, 24 rue Lhomond, 75231 Paris Cedex 05, France
	       }
 	 \date{Received / Accepted}

	  \authorrunning{G. Busquet \et}
 	 \titlerunning{The CHESS survey of the L1157-B1 bow-shock: high and low excitation water vapor}

\abstract{Molecular outflows powered by young protostars strongly affect the kinematics and chemistry of the natal molecular cloud through strong shocks resulting in substantial modifications of the abundance of several species. In particular, water is a powerful tracer of shocked material due its sensitivity to both physical conditions and chemical processes.}{As part of the ``Chemical Herschel Surveys of Star forming regions'' (CHESS) guaranteed time key program,
we aim at investigating the physical and chemical conditions of \hho\ in the brightest shock region B1 of the L1157 molecular outflow.}{We observed several ortho- and para-\hho\ transitions using HIFI and PACS instruments
on board \textit{Herschel} toward L1157-B1, providing a detailed picture of the kinematics and spatial distribution of the gas. We performed a Large Velocity Gradient (LVG) analysis to derive the physical
conditions of \hho\ shocked material, and ultimately obtain its abundance.
}{We detected 13 \hho\ lines with both instruments probing a wide range of excitation conditions. This is the largest data set of water lines observed in a protostellar shock that provide both the kinematics and the spatial information of the emitting gas.
PACS maps reveal that \hho\
traces weak and extended emission associated with the outflow identified also with HIFI in the o-\hho\ line at 556.9~GHz, and a compact ($\sim$10$''$) bright, higher-excitation region. The LVG analysis of \hho\ lines in the bow-shock show the presence of two gas components with different excitation conditions: a warm (\Tkin$\simeq$200-300~K) and dense ($n$(\hh)$\simeq$(1--3)$\times10^6$~\cmt) component with an assumed extent of 10$''$ 
and a compact ($\sim$2$''$-5$''$) and hot, tenuous (\Tkin$\simeq$900-1400~K, $n$(\hh)$\simeq$10$^{3-4}$~\cmt) gas component, which is needed to account for the line fluxes of high $E_{\rm u}$ transitions. The fractional abundance of the warm and hot \hho\ gas components is estimated to be (0.7--2)$\times10^{-6}$ and (1--3)$\times10^{-4}$, respectively. 
Finally, we identified an additional component in absorption in the HIFI spectra of \hho\ lines connecting with the ground state level. This absorption probably arises from the photodesorption of icy mantles of a water-enriched layer at the edges of the cloud, driven by the external UV illumination of the interstellar radiation field.}
{}{}


\maketitle

\section{Introduction}

Molecular outflows are among the most conspicuous manifestation of a nascent star. These outflows are known to result from the entrainment of circumstellar gas, swept-up by the primary jet, where a shock front is generated as a consequence of the supersonic impact of the jet with the natal cloud. Shocks heat, accelerate and compress
the ambient gas material switching on a complex chemistry that leads to an enhancement of the abundace of several species in the so-called ``chemically active outflows''  \citep[\eg][]{bachiller1996}.
The nature and properties of these shocks are still not well understood, in particular the role of the magnetic field. Water is predicted to be one of the main gas cooling agents in magnetized shocks, along with \hh\ and CO \citep[\eg][]{draine1983,kaufman1996,flower2010}. Thanks to its rich emission spectrum, transitions spanning a wide range of excitation conditions, and its sensitivity to local conditions \citep[\eg][]{bergin1998,vandishoeck2011}, \hho\ constitutes a powerful probe of the physics and chemistry of the shock outflow interaction. In particular, in shocked regions \hho\ abundance can increase by several orders of magnitude, up to $\sim$10$^{-4}$, through sputtering of grains mantles and formation in the gas phase at high temperatures \citep{hollenbach1989,kaufman1996,flower2010}.

\begin{figure}[!t]
\begin{center}
\begin{tabular}{c}
  	\epsfig{file=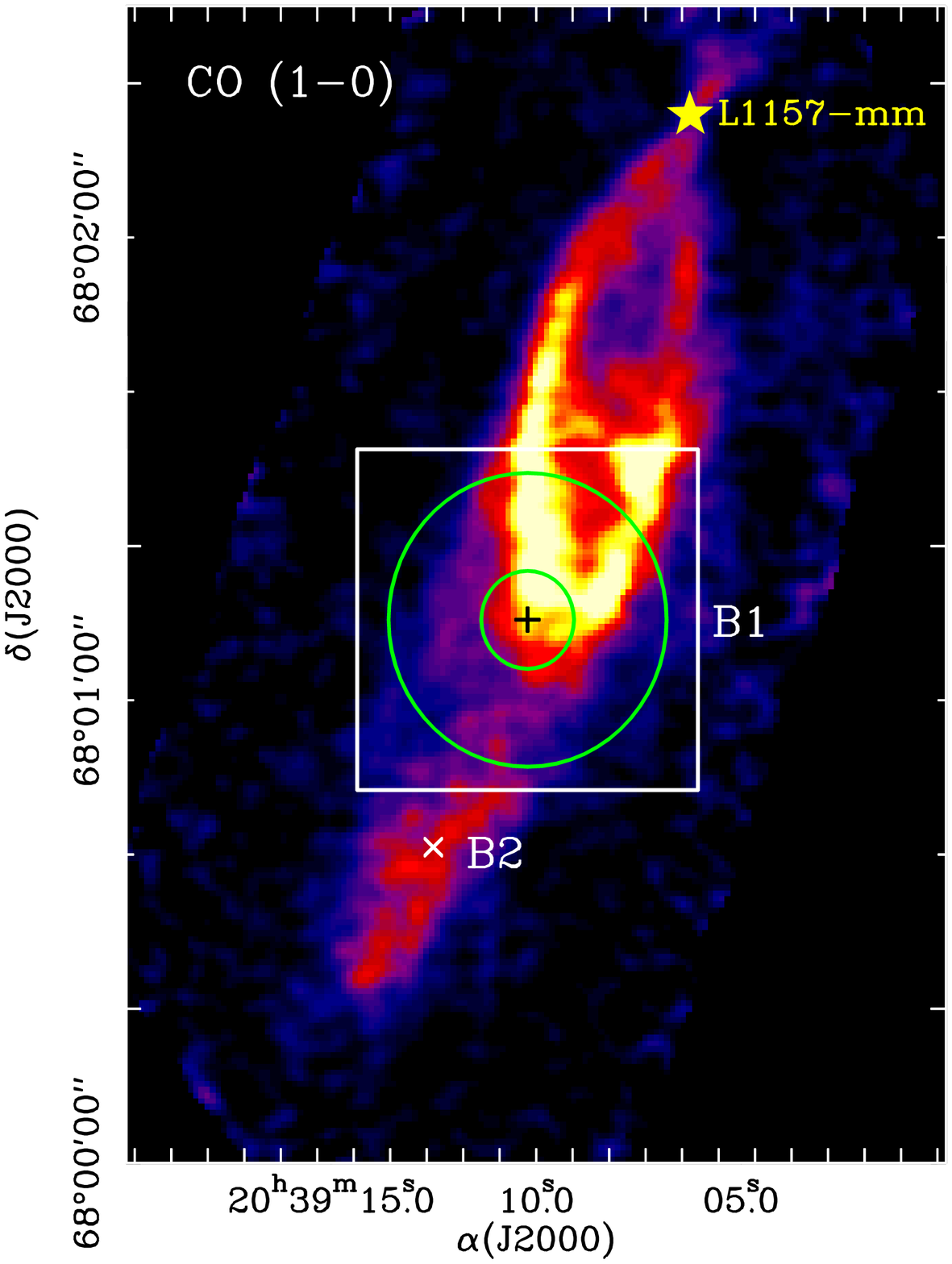,scale=0.4}
    \end{tabular}
     \caption{Southern blueshifted outflow lobe observed in CO\,(1--0) with the Plateau de Bure Interferometer (PdBI) by
     \citet{gueth1996}. The black cross marks the nominal position of the bow-shock L1157-B1 while the tilted white cross indicates the nominal position of the B2 shock \citep{bachiller1997}. The largest ($\sim$38$''$) and smallest ($\sim$12$\farcs7$) HIFI main-beams are indicated with circles. The field of view of the PACS observations is displayed with a white box. The star depicts the position of the protostar L1157-mm.}
\label{l1157bluelobe}
\end{center}
\end{figure}

The outflow powered by the low-mass Class\,0 protostar L1157-mm ($d\simeq$250~pc; \citealt{looney2007}) displays a rich specific chemistry which makes it the prototype of  ``chemically active'' outflows \citep[\eg][]{bachiller1997,bachiller2001,arce2008}. As such, it is an excellent laboratory to investigate the physical conditions and the formation routes of \hho\ and its role in the cooling of a typical protostellar outflow.
 The L1157 outflow has been studied in detail for more than two decades through many molecular lines and in a wide range of wavelengths, from the near-infrared \citep[\eg][]{davis1995,neufeld2009,nisini2010b} to the radio domain \citep[\eg][]{zhang2000,bachiller2001,tafalla1995}. Several compact shocked regions are found along both the blue- and red-shifted lobes
\citep[see \eg][]{gueth1998,nisini2007,nisini2010b}. In particular, the southern blueshifted lobe, shown in Fig.~\ref{l1157bluelobe}, consists of two limb-brigthened cavities each of them associated with a bow-shock, likely created by episodic events in a precessing jet \citep{gueth1996}.

Water emission in L1157-B1 was first detected with ISO by  \citet{giannini2001}; however, only three lines were detected and the physical conditions of \hho\ could not be constrained. Later on, {\em Odin} and SWAS observed the fundamental o-\hho\ line emission in the direction of the southern blueshifted lobe of the outflow \citep{bjerkeli2009,franklin2008}. The low angular resolution gave access only to properties averaged over the  entire outflow lobe. Assuming that both the \hho\ and the low-$J$ CO line emission originate in the same gas, these authors inferred an o-\hho\ abundance ranging between $10^{-6}$ and $2\times10^{-4}$. As part of the ``Water In Star-forming regions with Herschel'' (WISH) key program, \citet{nisini2010a} used PACS to map the o-\hho\ 179~\mum\ line over the entire outflow structure. These authors detected extended emission, with  several strong peaks associated with shocked knots, well spatially correlated with \hh\ rotational lines \citep{nisini2010b}.

The molecular bright shock region B1, in the southern lobe of the outflow (see Fig.~\ref{l1157bluelobe}), was selected as one of the targets of the key program ``Chemical HErschel Surveys of Star forming regions"  (CHESS\footnote{http://www-laog.obs.ujf-grenoble.fr/heberges/chess}) dedicated to unbiased spectral line surveys of prototypical star-forming regions \citep{ceccarelli2010} in the guaranteed time of the {\em Herschel} Space Observatory \citep{pilbratt2010}.

The CHESS survey of L1157-B1 offers a comprehensive view  on the water line emission in a typical protostellar bow-shock, considered as the benchmark for shock models \citep{gusdorf2008a,gusdorf2008,flower2010,flower2012}. A grand total of 13 water lines (both ortho and para) have been detected across the submillimeter and far-infrared window with the PACS spectro-imager \citep{poglitsch2010} and the  HIFI heterodyne instrument \citep{degraauw2010}, the largest data set of water lines detected so far in a protostellar shock.
Both instruments provide us with a detailed picture on the kinematics and the spatial distribution of the water emission in L1157-B1, allowing us to derive strong constraints on the water abundance and the physical conditions in the emitting gas.
The paper layout is as follows. In Sect.~2 we
summarize our observations. In Sect.~3 we present the main results of HIFI and PACS 
and in Sect.~4 we analyze the excitation conditions of \hho\ using a Large Velocity Gradient model and discuss the origin of the water emission in L1157-B1, presenting, for the first time, a detailed picture of the bow-shock structure through \textit{Herschel} observations of water lines.
Finally, in Sect.~5 we list the main conclusions.

\section{Observations and data reduction}

\begin{table*}[!ht]
\caption{List of \hho\ transitions$^{\mathrm{a}}$ observed with HIFI in L1157--B1. Peak intensity (in $T_{\mathrm{mb}}$ scale), peak velocity, and integrated intensity between $-$40 and $+$2.6~\kms\ are reported. The uncertainties are indicated in parenthesis.}
\begin{center}
\begin{tabular}{lcccccccccc}
\hline\hline
Transition 	& Frequency 	&$\lambda$	&$E_{\mathrm{u}}/k_{\rm{B}}$ &HPBW & $\eta_{\rm{mb}}$  & Obs$\_$Id  &rms$^{\mathrm{b}}$ &$T_{\mathrm{peak}}$ &\velo$_{\mathrm{peak}}$  &$\int\,T_{\mathrm{mb}}\,d$\velo  \\
			 &(GHz)     	&(\mum)		&(K)   		&(arcsec)  &		  &   & (mK)  &(K) &(\kms)  &(K~\kms)	\\
\hline
o-\hho\,$1_{10}-1_{01}$   		&\phn556.936 	&538.66		&\phn26.7  &38.1    &0.75	 & 1342181160   &\phnn8\phs  &1.04(0.16) &$-1.9(0.3)$ &13.0(0.1)   \\
\phwater\,$3_{12}-3_{03}$ 	&1097.365    	&273.38		&215.1	 &19.3	    &0.74	   &  1342196453  &\phn26\phs  &0.29(0.04) &$-2.0$(0.3)  	 &\phn4.1(0.2)	\\ 	
\phwater\,$3_{12}-2_{21}$ 	&1153.127    	&260.17		&215.1   &18.3	    &0.64	& 1342207691    &\phn39\phs  &0.18(0.03) &	$-1.9$(0.1)  		&\phn1.3(0.3) \\
\phwater\,$3_{21}-3_{12}$ 	&1162.912    	&257.98		&271.0	&18.3       &0.64	& 1342207691    &\phn36$^{\mathrm{c}}$ &0.11(0.03) &$-1.7(0.4)$   &\phn0.5(0.2) \\
\phwater\,$2_{21}-2_{12}$ 	&1661.008    	&180.49		&159.8 	 &12.7	    &0.71   & 1342196538     &138\phs 	&0.36(0.14)	&$-1.7(0.1)$	 	&\phn2.0(0.8)\\
\phwater\,$2_{12}-1_{01}$  	&1669.905    	&179.52		&\phn80.1     &12.7       &0.71  & 1342207689    &\phn56\phs &0.94(0.14) &$-5.5$(0.3)   &12.7(0.4)\\
p-\hho\,$2_{11}-2_{02}$   		&\phn752.033 	&398.92		&136.9  &28.2       &0.75	&  1342207611  &\phn20\phs 	&0.46(0.07)  &$-3.6(0.3)$	 &\phn4.6(0.1)		\\
\phwater\,$2_{02}-1_{11}$ 	&\phn987.927 	&303.67		&100.8   &21.5	    &0.74  &  1342207640  &\phn50\phs	&0.79(0.12) &$-$3.3(0.2)		&11.0(0.3)	\\
\phwater\,$1_{11}-0_{00}$        &1113.343    	&269.47		&\phn53.4   &19.1       &0.74 &  1342207388  &\phn56\phs	 &0.79(0.12) &$-4.1$(0.1)   &11.1(0.4)	\\
\hline
\end{tabular}
\tablefoot{
\tablefoottext{a}{Frequencies taken from the spectroscopic catalog JPL \citep{pickett1998}.}
\tablefoottext{b}{Root mean square (rms) noise are given for an interval of 0.5~MHz.}
\tablefoottext{c}{Root mean square (rms) noise is given for an interval of 1.5~MHz.}
}
\end{center}
\label{hifilines}
\end{table*}

\begin{table*}[!t]
\caption{List of \hho\ transitions$^{\mathrm{a}}$ detected with PACS in L1157--B1, at the nominal position of B1 (spaxel centered at offset ($0'', 0''$)) and at the high-$J$ CO peak (spaxel at offset ($-5'', 7'')$).}
\begin{center}
\begin{tabular}{lccccccc}
\hline\hline
Transition 		  &Frequency 	&$\lambda$	&$E_{\mathrm{u}}/k_{\rm{B}}$  &HPBW    &\multicolumn{2}{c}{Flux$^{\rm{b}}$}   	&Line$\_$Id  \\
\cline{6-7}
			  &(GHz)     	&(\mum)		&(K)   		              &(arcsec)  &\multicolumn{2}{c}{($\times10^{-17}$\,W\,m$^{-2}$/pixel)}	&	\\
			  &		&		&			      &		 &offset (0$'',0''$)    &offset $(-5'',7'')$ \\
\hline
o-\hho\,$2_{21}-2_{12}$    &1661.008    &180.49		&159.8                       &12.7	 &\phn2.6$\pm$0.2	&\phn2.7$\pm$0.3		&1 \\	
\phwater\,$2_{12}-1_{01}$   &1669.905    &179.52        &\phn80.1                    &12.7       &12.1$\pm0.2$          &12.9$\pm$0.3 		&2\\
\phwater\,$3_{03}-2_{12}$  &1716.769    &174.63         &162.5                       &12.3       &\phn6.2$\pm0.3$       &\phn6.5$\pm$0.3	&3	\\
\phwater\,$4_{23}-4_{14}$  &2264.149    &132.41         &397.9			     &\phn9.4    &$<$0.6$^{\rm{c}}$	&\phn0.6$\pm$0.2 	&4\\
\phwater\,$4_{14}-3_{03}$\,$^{\mathrm{d}}$ &2640.474    &113.54         &289.2                         &\phn8.0     &\phn2.8$\pm$0.5     &\phn3.7$\pm0.4$	&5 \\
\phwater\,$2_{21}-1_{10}$   &2773.978    &108.07         &159.8			     &\phn7.6	 &\phn3.2$\pm$0.4	&\phn3.9$\pm$0.5 		&6\\
\phwater\,$3_{21}-2_{12}$   &3977.046    &\phn75.38      &271.0			     &\phn5.3    &$<$2.1$^{\rm{c}}$	&\phn1.9$\pm$0.7 	&7\\
p-\hho\,$3_{13}-2_{02}$   &2164.132 	&138.53		&204.7               	     &\phn9.8    &\phn2.7$\pm$0.3 	&\phn2.7$\pm$0.3		&8\\
\phwater\,$4_{04}-3_{13}$ &2391.573 	&125.35		&319.5               	     &\phn8.9  	 &\phn1.1$\pm$0.3       &\phn1.4$\pm$0.3		&9\\
\hline
\end{tabular}
\tablefoot{
\tablefoottext{a}{Frequencies taken from the spectroscopic catalog JPL \citep{pickett1998}.}
\tablefoottext{b}{Line fluxes obtained with a Gaussin fit of the line profile.}
\tablefoottext{c}{$3\sigma$ upper limit based on the rms estimated at offset ($-5'',7''$).}
\tablefoottext{d}{This line is blended with the CO\,(23--22) transition. To estimate the flux of the water line we subtracted the flux of CO\,(23--22) transition
predicted by the model presented in \citet{benedettini2012}.}
}
\end{center}
\label{tpacsflux}
\end{table*}

\subsection{HIFI observations}

The HIFI observations were performed in double beam switching mode during 2010 towards the nominal position of B1: $\alpha(J2000)$=$20^{\mathrm{h}}39^{\mathrm{m}}10\fs2$, $\delta(J2000)$=68$\degr01\arcmin10\farcs5$.
Both polarizations (H and V) were observed simultaneously. The receiver was tuned in double sideband (DSB).
Most of the submillimeter window was covered in an unbiased way with HIFI, and the observations were carried out in spectral scanning mode. In order to study the properties of the \hho\ gas in the high-velocity wings of the outflow, a few lines were observed in pointed mode in order to reach an excellent signal-to-noise ratio (SNR). 

We used the Wide Band Spectrometer (WBS), which provides a frequency resolution of 0.5~MHz (\ie\ velocity resolution between 0.1~\kms\ and 0.4~\kms, depending on the wavelength).
The data were processed with the ESA-supported package \textit{Herschel Interactive Processing Environment}\footnote{HIPE is a joiny development by
the \textit{Herschel} Science Ground Segment Consortium, consisting of ESA, the NASA \textit{Herschel} Science Center, and the HIFI, PACS and SPIRE consortia} (HIPE, \citealt{ott2010}) version 6 package. After level 2 fits files were exported and transformed into GILDAS\footnote{See http://www.iram.fr/IRAMFR/GILDAS} format for baseline subtraction and subsequent sideband deconvolution, which was performed manually. The relative calibration between both receivers (H and V) was found very good, and both signals were co-added in order to
improve the noise rms of the data.

The spectral resolution was then degraded to a common velocity resolution of 1~\kms\ in the final single side band (SSB) data set. Uncertainty in the flux calibration were estimated to be $\sim$20\,\%. In Table~\ref{hifilines} we summarize the observational
parameters of the \hho\ transitions detected with HIFI (frequency, wavelength, upper level energy).
The main-beam intensity, peak velocity, and the integrated intensity for each transition are also reported. Intensities are expressed in units of main-beam brightness temperature.  The telescope
parameters (half power beamwidth (HPBW) and main-beam efficiency ($\eta_{\rm{mb}}$)) are taken from \citet{roelfsema2012}.

\begin{figure*}[!t]
\begin{center}
\begin{tabular}{c}
  	\epsfig{file=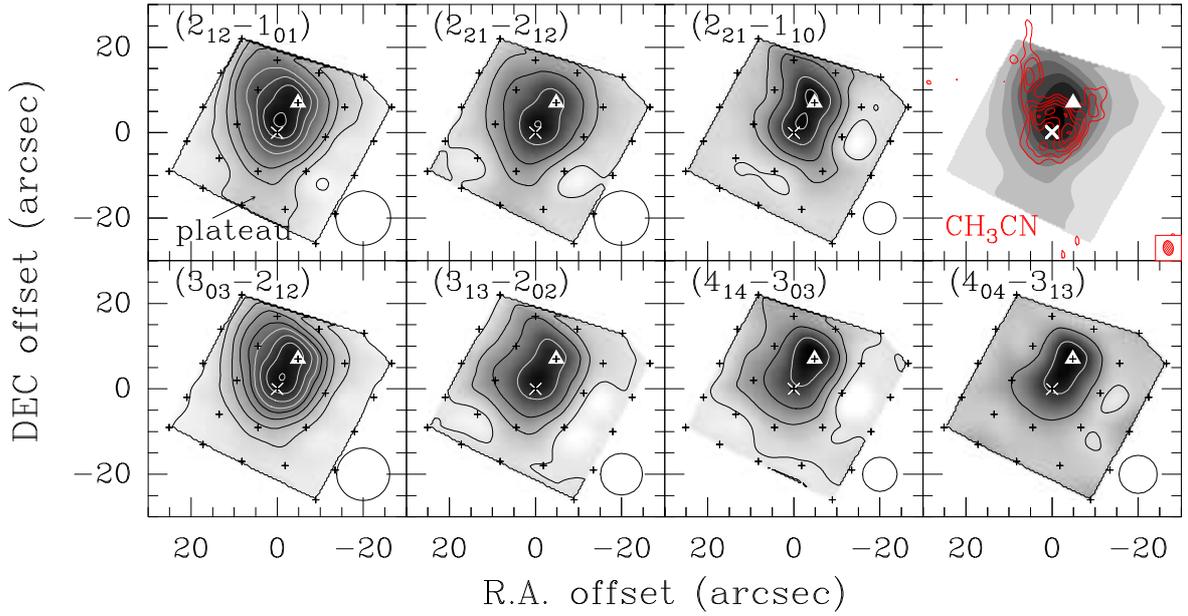,scale=0.98}
    \end{tabular}
     \caption{PACS maps of \hho\ line fluxes (increasing upper level energy from top left to bottom right). First contour corresponds to $1\sigma$ flux level of each transition.
Contours step is $2\sigma$ in all lines except for the o-\hho\,($2_{12}-1_{01}$) and o-\hho\,(3$_{03}-2_{12}$), for which contours step is $6\sigma$ and $3\sigma$, respectively, where $\sigma$ is listed in Table~\ref{tpacsflux}. The top right panel shows an overlay of the o-\hho\,($2_{12}-1_{01}$) map (grey scale) with the PdBI \chtcn\,(8--7) $K$=0--2 image (red contours) from \citet{codella2009} tracing the bow-shock. The synthesized beam of PdBI, $3\farcs4\times2\farcs3$ (P.~A.=10$\degr$), is shown in the bottom right corner of this panel. Crosses mark the central position of each spaxel of the PACS field of view. The white triangle indicates the position of CO peak traced by high-$J_{\mathrm{up}}$ PACS lines \citep{benedettini2012}
and the tilted cross depicts the nominal position of the B1 shock. The Half Power Beam Width (HPBW) of each transition is indicated in the bottom right corner of each panel and listed in Table~\ref{tpacsflux}.}
\label{owatermaps}
\end{center}
\end{figure*}

\begin{figure*}[!ht]
\begin{center}
\begin{tabular}{c}
    \epsfig{file=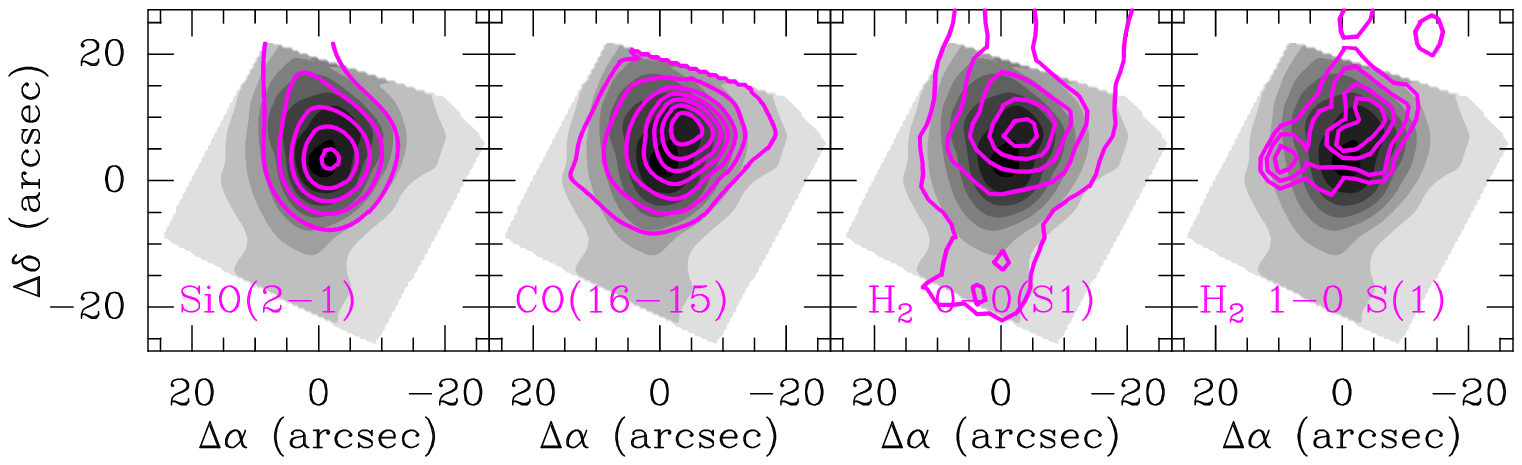,scale=1,angle=0}\\

     \end{tabular}
     \caption{Contour maps of the SiO\,(2--1) high-velocity (\velo\,<--8~\kms) emission from Gueth \et\ (1998) convolved to the $12\farcs7$ PACS resolution, the CO\,(16--15) line observed with PACS \citep{benedettini2012}, \hh\,0--0 S(1) from \citet{nisini2010b}, and \hh\,1--0 S(1) from \citet{caratti2006}, overlaid on the \hho\,($2_{12}-1_{02}$) map at 179~\mum\  (grey scale).  In each panel the transition in contours is indicated.}
\label{watercosio}
\end{center}
\end{figure*}

\subsection{PACS observations}

The PACS observations were carried out on May 25th, 2010 in line spectroscopy mode in order to obtain a full range spectrum of the molecular line emission towards B1, from 55-95.2~\mum\ and from 101.2-210~\mum. The spectral scan was centered at the nominal position of B1 (see above) and produced a single $5\times5$ spectral
map of $9\farcs4$ square spatial pixels (hereafter spaxels) over a $47''\times47''$ field of view.
Two observations were conducted for the 161.5-190.2~\mum\ range. Both measurements are in good agreement, with a discrepancy
at the most $\sim$10\,\%.
The resolving power ranges from 1000 to 4000 (\ie\ spectral resolution of $\sim$75-300~\kms)
depending on the wavelength, hence water lines are unresolved. PACS data were processed with HIPE version 5.0.
 The absolute flux scale was determined from observations of Neptune by normalizing the observed flux to the telescope background, with an
estimated uncertainty of $\sim$10\,\% for  $\lambda$<190~\mum\ (\ie\  where all the water PACS lines lie). Further details of the PACS observations are described in \citet{benedettini2012}, where the emission lines of CO, OH, and [OI] lines are presented and discussed.

In Table~\ref{tpacsflux} we list the detected transitions, giving their frequency, wavelength, upper energy level, and beam size.
We extracted the flux toward each spaxel adopting a Gaussian instrumental response.
The line fluxes measured toward the two brightest spaxels, at offsets ($0''$, 0$'')$ and ($-5''$, 7$'')$, associated with the nominal position of B1 and the high-excitation CO emission peak \citep{benedettini2012}, respectively, are reported in Table~\ref{tpacsflux}.

\subsection{Cross calibration}

Four lines were observed by both PACS and HIFI instruments, which allowed us to check the consistency of the calibration:
H$_2$O\,($2_{12}$--$1_{01}$) at 179~\mum, ($2_{21}$--$2_{12}$) at 180~\mum, CO\,(16--15), and CO\,(14--13).
Using the PACS maps, we estimated the line intensities in the HIFI main-beam solid-angle towards the nominal position of B1. Comparison of the HIFI- and PACS-based line intensities shows a very good agreement for H$_2$O\,($2_{12}-1_{01}$) and CO\,(14--13), where the integrated intensities are 13.7/14.2~K\,\kms\ and 5.0/5.5~K\,\kms, respectively (for HIFI/PACS instruments), resulting in a discrepancy of $\sim$10\,\%.
For the H$_2$O\,($2_{21}-2_{12}$)
and CO\,(16--15) lines, the HIFI/PACS integrated intensities are 2.0/1.8~K\,\kms\ and 3.4/2.7~K\,\kms, respectively. For these lines
the discrepancy is larger, about
$\sim$20\,\%, but always within the absolute flux calibration uncertainty.
We note that both lines are weaker, and the SNR
of the data much lower than the other two lines. Overall, we conclude that the agreement
between the HIFI and PACS calibration scales is very good.

\section{Results}

We have detected 13 H$_2$O transitions  with a flux  above the $5\sigma$ level: 7 H$_2$O transitions
(5 ortho, 2 para) in the PACS range (55--210~\mum),  8 H$_2$O transitions (5 ortho, 3 para)  with HIFI between
672~\mum\ and 180~\mum. Note that the 179~\mum\ and 180~\mum\ lines have been detected with both instruments. The o-\hho\,($3_{21}-3_{12}$) transition is
detected at a $3\sigma$ level with HIFI (see Table~\ref{hifilines}). Two additional o-H$_2$O transitions ($3_{21}-2_{10}$ at 75.38~\mum\ and $4_{23}-4_{14}$ at 132.41~\mum) are tentatively detected with PACS at the $3\sigma$ level at the offset position ($-5''$, 7$'')$ (see Table~\ref{tpacsflux}).
Overall, we detected only transitions of rather low $E_{\rm u}$, with values ranging between 26.7~K and 319.5~K.

\begin{figure}[!t]
\begin{center}
\begin{tabular}{c}
  	\epsfig{file=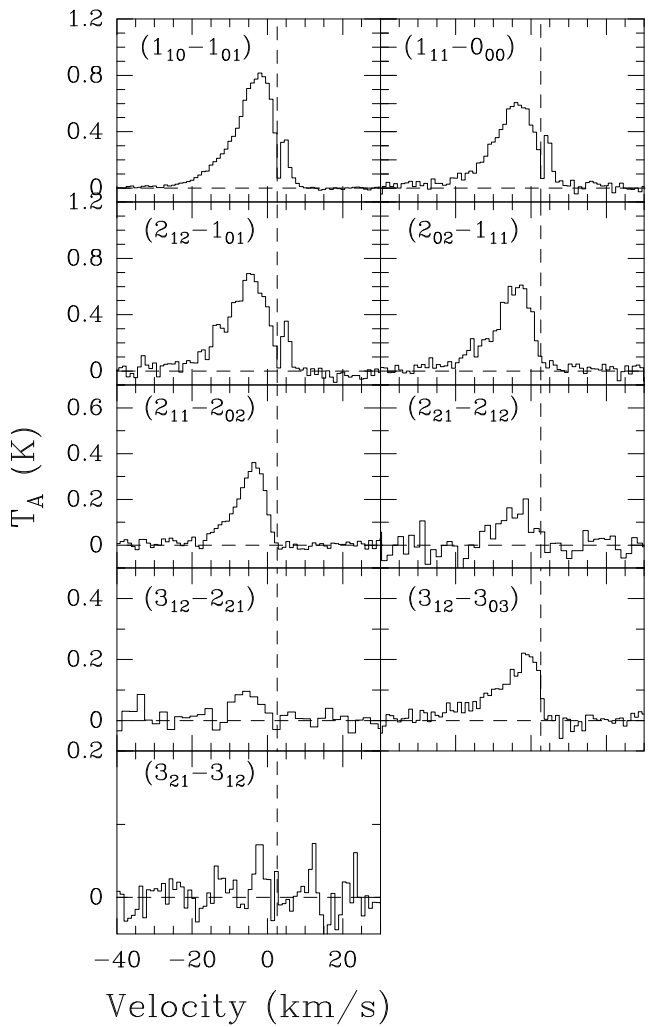,scale=0.8,angle=0}
        \end{tabular}
     \caption{HIFI \hho\ spectra of L1157-B1 smoothed to a velocity resolution of 1~\kms.
     The \hho\ transition
     is indicated in each panel. The vertical dashed line marks the ambient LSR velocity \vel$\sim$2.6~\kms\ from C$^{18}$O emission \citep{bachiller1997}.}
\label{waterprofiles}
\end{center}
\end{figure}

\subsection{\hho\ spatial distribution}

Maps of o-H$_2$O and p-H$_2$O lines observed with PACS are presented in Fig.~\ref{owatermaps}.
Water is detected over the entire outflow cavity,  both upstream and downstream of the bow-shock; its distribution overlaps rather well with the B1 bow-shock as traced by \chtcn\ \citep[][see top right panel]{codella2009} and the outflow walls of the B1 cavity, traced by CO at the PdBI \citep[][see Fig.~\ref{l1157bluelobe}]{gueth1996}. Downstream of B1, weak emission is present $20\arcsec$ away from the shock in several transitions, including the $3_{03}-2_{12}$ and $2_{12}-1_{01}$ lines, in agreement with
\citet{nisini2010a}.
This extended emission, which consists of a plateau of low \hho\ brightness, is related to the ouflow, possibly from the B2 outflow cavity.

Leaving aside the contribution of the plateau to the emission,  the distribution of the water emission in B1 displays little variation between the various transitions, with a typical deconvolved size at Full-Width at Half Maximum (FWHM) of $\sim$10$''$.
Overall, the emission appears elongated along the major outflow axis.  The H$_2$O brightness peak is located $\approx$6$''$ north of the center of the PACS array, lying approximately halfway between the nominal position of B1 and the high-$J$ CO emission peak identified by \citet{benedettini2012}, at the interface between spaxels (2,2) and (3,2) at offsets (0$'', 0''$) and ($-5'', 7''$), respectively. However, size and position determination from the PACS undersampled data are just indicative and they suffer large uncertainties.

It is interesting to compare the morphology of the $2_{12}-1_{01}$ line  with that of other shock tracers 
(see Fig.~\ref{watercosio}).
One can see that both \hho\ and SiO peak at the same position, between offset (0$'', 0''$) and ($-5'', 7''$). Similar to \hho, the emission of the mid-IR \hh\,0--0 S(1) pure-rotational line observed with {\em Spitzer} \citep{nisini2010b} is extended, partly tracing the B1 cavity, while CO\,(16--15) and the near-IR \hh\,1--0 S(1) ro-vibrational line \citep{caratti2006} are rather compact and peak at offset ($-5'', 7''$), coinciding with the partly dissociative shock driven by the impact of the jet against the B1 cavity \citep{benedettini2012}. However, it is worth noting that there is also bright  \hho\ emission at the peak of the high-$J$ CO position, suggesting that part of the \hho\ emission coincides with CO.
The good match between SiO\,(2--1) and \hho\,($2_{12}-1_{01}$), both in terms of spatial distribution and the line profiles in the high-velocity range \citep[see Fig.~2-b of][]{lefloch2012}, provides us with an estimate of the size of the water line emission,  $\approx$10$''$, consistent with the PACS determination.

\subsection{Line profiles}

Figure~\ref{waterprofiles} shows a montage of the water line spectra observed with HIFI smoothed to a  velocity resolution of 1~\kms. Water line profiles are rather broad, with a FWHM$\sim$10~\kms. The bulk of the emission in all transitions is clearly {\em blueshifted} with respect to the cloud systemic velocity \vel=+2.6~\kms\ \citep{bachiller1997}. For lines with a SNR
high enough, \eg\ $3_{12}$--$3_{03}$, $3\sigma$ emission is detected at velocities up to $-$30~\kms.

The coexistence of multiple excitation components in L1157-B1 has been studied recently by \citet{lefloch2012}, who showed, based on the spectral slope, that the CO line emission arises from three different emitting regions. These components were tentatively identified as the jet impact shock region ($g_{\rm{1}}$),  the cavity walls of the L1157-B1 bow-shock ($g_{\rm{2}}$), and the cavity walls from the earlier ejection episode that produced the L1157-B2 bow-shock ($g_{\rm{3}}$). An schematic view of all these components is presented in Fig.~\ref{cartoon}. The authors showed that  each component is characterized by an specific excitation temperature.
We found that the profile of the o-\hho\,($3_{12}-3_{03}$) transition follows the same specific spectral signature observed for the CO\,(16--15) line profile \citep[see Fig.~2-b of][]{lefloch2012}, which is defining the $g_{\rm{1}}$ component.

\begin{figure}[!t]
\begin{center}
\begin{tabular}{c}
  	\epsfig{file=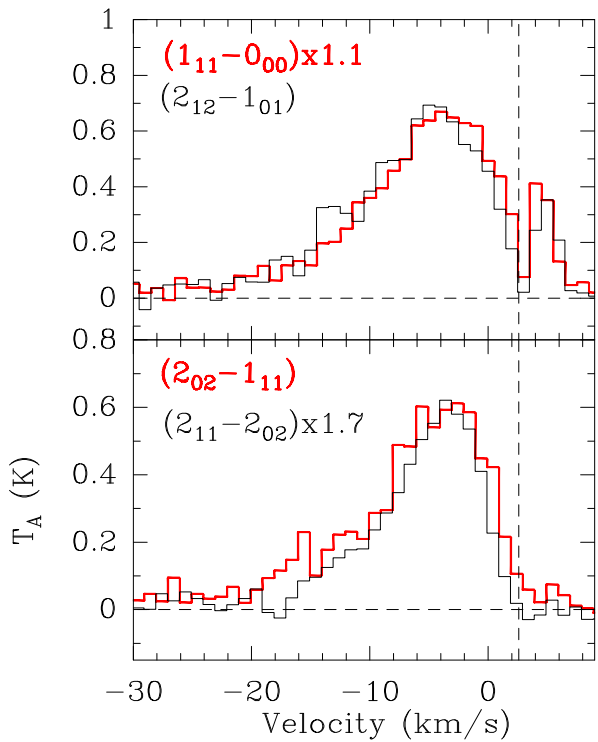,scale=0.7,angle=0}
        \end{tabular}
     \caption{Comparison of HIFI \hho\,($2_{12}-1_{01}$) and ($1_{11}-0_{00}$) lines ({\em top}) at 1669~GHz and 1113~GHz, respectively, and between the \hho\,($2_{02}-1_{11}$) at 988~GHz and the ($2_{11}-2_{02}$) at 752~GHz lines ({\em bottom}).
     The \hho\ transitions are labeled in each panel. The vertical dashed line marks the ambient LSR velocity \vel$\sim$2.6~\kms\
     \citep{bachiller1997}.}
\label{hificomp}
\end{center}
\end{figure}

Despite the different beam sizes of the HIFI lines, Figure~\ref{hificomp}
shows a good
match between the profiles of the water lines $2_{11}-2_{02}$ at 752~GHz and $2_{02}-1_{11}$ at 988~GHz, and between $1_{11}-0_{00}$ at 1113~GHz and  $2_{12}-1_{01}$ at 1669~GHz, respectively. This defines two groups of water lines, each of them following a specific pattern suggesting that the lines within each group arise from the same region. Whereas the lines at 1669~GHz and 1113~GHz both peak at
$-$5~\kms, the lines at 752~GHz and 988~GHz peak at
$-$3~\kms.

A narrow dip ($\Delta$\velo=1.4~\kms) is observed at the systemic cloud velocity \vel\,=2.6~\kms\ in the spectra of the three transitions that connect with the ground state level
resulting in a double-peak profile (see Fig.~\ref{waterprofiles} and Sect.~4.3 for further details).

Weak redshifted emission is detected in these transitions {\em only},  up to velocities of +10~\kms. \citet{bjerkeli2013} showed that this weak redshifted emission is extended all over the southern outflow lobe. It is worth noting that this redshifted emission is also detected in the low-$J$ HCN and
\hcop\ lines \citep{bachiller2001,benedettini2007}. The lack of emission in other tracers such as CS, \chtoh, H$_2$CO, or SiO suggests that this component has different excitation conditions from the main, blueshifted outflow component. The redshifted emission most likely arises from material located on the rear side of the cavity.

\begin{figure}[!t]
\begin{center}
\begin{tabular}{c}
	\epsfig{file=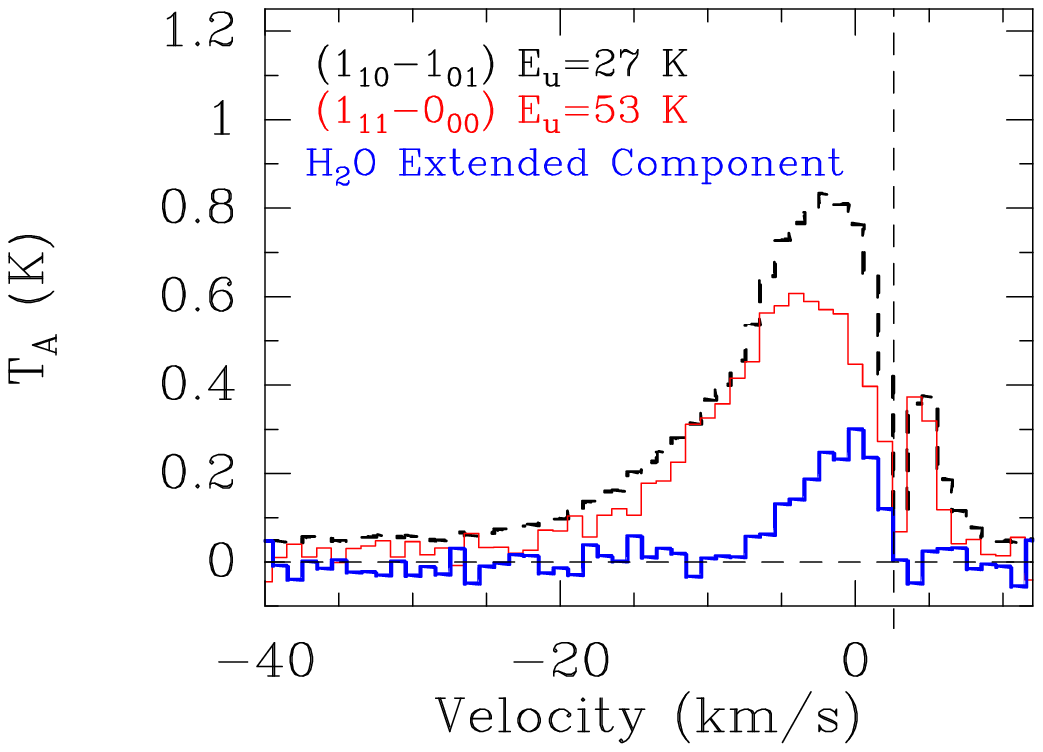,scale=0.8,angle=0}\\
          \epsfig{file=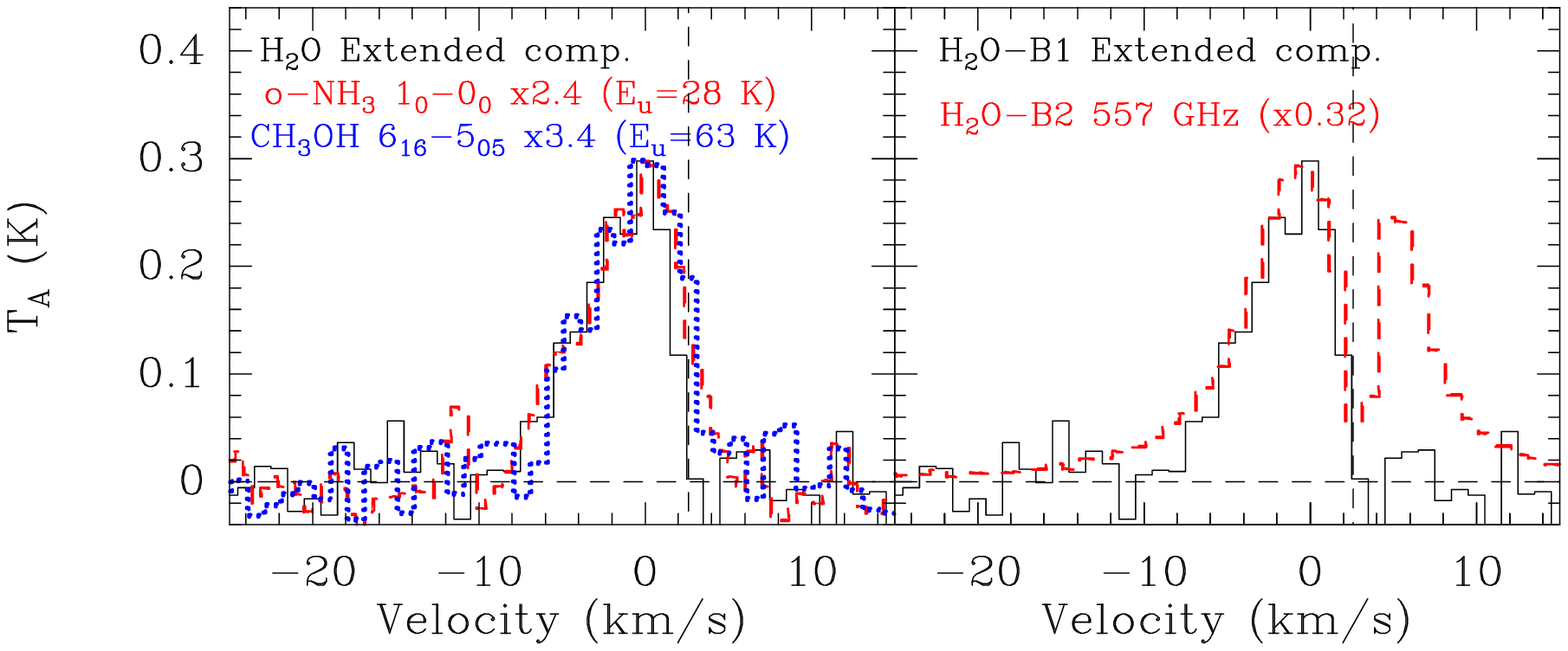,scale=0.46,angle=0}\\
        \end{tabular}
     \caption{\emph{Top:} Comparison of HIFI spectra for \hho\ transitions connecting with the ground state:  o-\hho\,($1_{10}-1_{01}$) shown by the dashed black line
     and p-\hho\,($1_{11}-0_{00}$) shown by the thin red solid line.
     The spectrum shown in blue (solid thick line) is the residual emission after subtracting the emission of  p-\hho\,($1_{11}-0_{00}$) line from the o-\hho\,($1_{10}-1_{01}$), referred to as extended component. \emph{Bottom:} Comparison of the extended component seen in the \hho\,($1_{10}-1_{01}$) line (black solid line) with the o-\nh\,(1$_{0}-0_{0}$) (red dashed line) and the \chtoh\,($6_{16}-5_{05}$) (blue dotted line) spectra obtained from \citet{codella2010} (left panel).  Comparison of the o-\hho\,($1_{10}-1_{01}$) line at 556.9~GHz observed in B1 (extended component; black solid line) and in L1157-B2  (red dashed line) from \citet{vasta2012}. In all panels the vertical dashed line marks the ambient LSR velocity \vel$\sim$2.6~\kms\ \citep{bachiller1997}.}
\label{waterextdcompare}
\end{center}
\end{figure}

\subsection{Outflow emission}

Figure~\ref{waterextdcompare} (top panel) shows that the profile of the  o-\hho\,($1_{10}$--$1_{01}$) line at 556.9~GHz presents a significant excess of emission at low velocities  compared with the two other lines connecting the ground state level (the $2_{12}$--$1_{01}$ at 1669~GHz and the $1_{11}-0_{00}$ at 1113~GHz transitions) that display similar line profiles (see Fig.~\ref{hificomp}).
This is a priori surprising as the lines at 1113~GHz and 1669~GHz are sensitive to somewhat different excitation conditions.

We subtracted the profile of the p-\hho\,($1_{11}-0_{00}$) line from the o-\hho\,($1_{10}-1_{01}$) line and show its residual emission in Fig.~\ref{waterextdcompare} (thick spectrum in the top panel). This residual emission has a peak intensity of
$T_{\mathrm{mb}}$$\simeq$0.38~K and it peaks at \velo=0~\kms. This component spans a relatively narrow range of velocities, from $\sim$2.6~\kms\ up to $-$8~\kms, and its integrated intensity, in $T_{\mathrm{mb}}$ scale,  is 2.3~K\,\kms.

\begin{figure*}[!t]
\begin{center}
\begin{tabular}{c}
          \epsfig{file=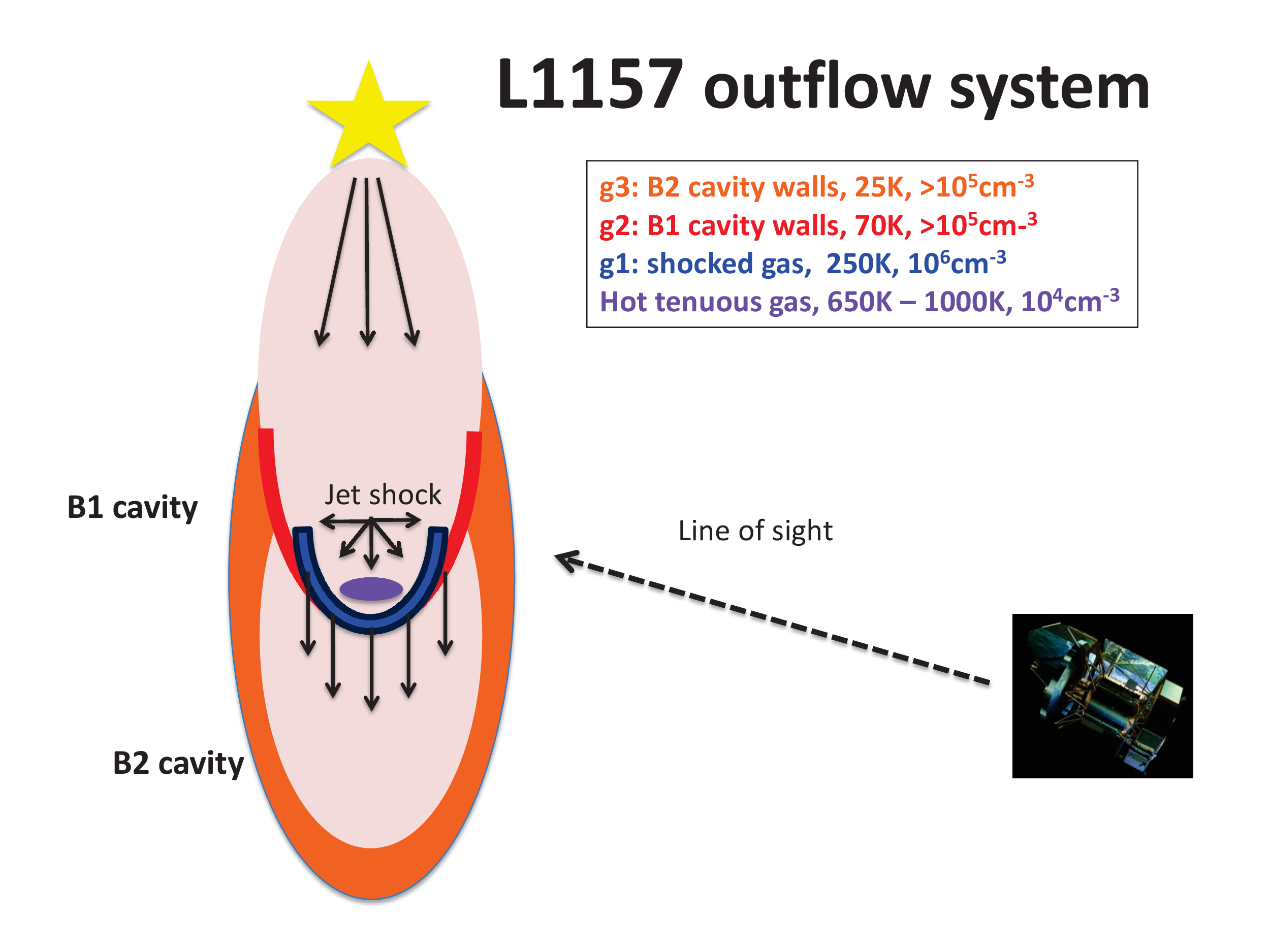,scale=0.55} \\
  	   \end{tabular}
    \caption{Sketch of the L1157 blue-lobe outflow system. The B1 and B2 outflow cavities are indicated in red and orange, respectively. The two shock components identified through \hho\ lines are displayed in blue (warm shocked gas) and in light violet for the hot tenuous gas. In the top right corner of the image we report on the physical conditions of each component. The observer, represented by the {\em Herschel} satellite, is indicated to the right side of the image.}
\label{cartoon}
\end{center}
\end{figure*}

As can be seen in Fig.~\ref{l1157bluelobe}, the HIFI beam at 1113~GHz and 1669~GHz collects emission from a region at the apex of the bow-shock, with a typical size of  $\approx$10$\arcsec$ (see also \citealt{lefloch2012}) whereas the 556.9~GHz actually collects also emission associated with the B1 cavity walls and the entrained gas, downstream and eastward of the B1 cavity, associated with the B2 ejection.
An additional clue on the origin of the extended component is obtained by comparing the 556.9~GHz line profiles of the extended component and the older outflow cavity L1157-B2
 (see also Fig.~\ref{l1157bluelobe}) observed by \citet{vasta2012}. As can be seen in Fig.~\ref{waterextdcompare} (bottom right panel),
both profiles show an excellent match at blueshifted velocities, suggesting a common origin. Interestingly, an excellent match was observed in the CO $J=$3--2 profiles of the $g_{\rm 3}$ component and the L1157-B2 shock by \citet{lefloch2012}.
Comparison of the line profile of the extended \hho\ component with the o-\nh\,($1_{0}-0_{0}$) and \chtoh\,($6_{16}-5_{05}$) lines \citep{codella2010}, observed at a similar angular resolution with HIFI ($\sim$38$''$), reveals a very good agreement, supporting the hypothesis that they have a common origin and all trace the same gas.
We propose that the broad HIFI beam at 556.9~GHz is actually tracing an extended component, of low excitation, for which the beam of HIFI  is less sensitive to  at the frequency of the  \hho\  lines at 1113~GHz (1669~GHz), as its  size decreases from 38\arcsec to $19\farcs1$ ($12\farcs7$).  Such a component could represent the counterpart at 556.9~GHz of the plateau evidenced by PACS, south of B1. Lack of angular resolution prevents from being more specific about the origin of the extended component, and a comparison with the 556.9~GHz map presented by \citet{bjerkeli2013} would help to support our interpretation.

The \hho\ emission from  the outflow has been recently analyzed by \citet{bjerkeli2013}, who presented a detailed study of the physical properties (molecular mass, dynamical time-scale, momentum, kinetic energy, etc) in the outflow using CO and \hho\ lines. In what follows, we will concentrate
on the physical conditions in the bow-shock, where the bulk of \hho\ emission originates from.

\section{Analysis and discussion}

We have determined the physical conditions of the B1 shock region
from modelling the water line emission using a radiative transfer code in the Large Velocity Gradient (LVG) approximation.
It is worth noting that the B1 shock position is about $1\farcm5$ far away from the protostar L1157-mm; the continuum emission detected in the submm/far-IR range is faint enough that infrared radiative pumping of the \hho\ lines can be neglected.

We first present our approach to the modelling of the L1157-B1 emission (Sect.~4.1),  we then discuss  the best-fit solution to the emission from the B1 shock (Sect.~4.2), and we show its consistency with the previous works on CO and \hh. We assess the influence of various parameters of the modelling, in particular the actual value of the ortho to para \hh\ ratio in the shocked gas. Finally, we study the origin of the water absorbing layer in the cloud (Sect.~4.3), and report on the water abundance and far-IR cooling (Sect.~4.4).
For the sake of clarity, we display in Fig.~\ref{cartoon} the physical structure emerging from our \hho\ line study that summarizes the main results to be presented in this section. Briefly, we identified five components: the cloud seen in absorption, the outflow material from the B1 ($g_{\rm 2}$) and B2 ($g_{\rm 3}$) cavities, the jet impact shock region ($g_{\rm 1}$) and a compact hot gas component.

\subsection{Modelling}

Previous studies at millimeter and infrared wavelengths \citep[][and more recently \citealt{benedettini2013}]{benedettini2007,codella2009,takami2011}
reveal a complex density, temperature, and velocity 
structure, with several emission knots of shocked gas in various tracers.
In such complex environment, a comprehensive modelling of the water line emission from the bow-shock, notoriously a very difficult task, is just
a problem too difficult to handle if one consider the angular resolution of the data (at best comparable to the size of the region), the one-dimensional nature of the source geometrical modelling, and the radiative transfer code used. Our goal here is to identify the main shock components responsible for the \hho\ emission detected, and, within the uncertainties inherent to the calibration and geometry adopted, the physical conditions of these components.

The PACS maps (Fig.~\ref{owatermaps}) show that the \hho\ emission does not peak at the nominal position of B1, where the HIFI beam is centered; the larger HIFI beams encompass the emission peak while for the smaller HIFI beams the peak is partially covered (see Fig.~\ref{l1157bluelobe}). It is all the more important to carry out LVG calculations using water fluxes measured over the same source solid angle; we therefore convolved all the PACS maps to a common angular resolution of $12\farcs7$ to measure the flux towards the nominal position of B1. For the HIFI lines, we convolved the o-\hho\,($2_{12}-1_{01}$) PACS map at the resolution of the different HIFI beams. Assuming that all HIFI lines have
the same spatial distribution as the o-\hho\,($2_{12}-1_{01}$) line 
we can then derive the HIFI beam filling factor. As a test, we compared the beam filling factors obtained from PACS maps of the 108, 138, and 125~\mum\ lines. In practice, we obtain very similar correcting factors, which made us feel confident in the robustness of the procedure and the results obtained. The fluxes of all the \hho\ lines estimated in a beam of $12\farcs7$ are listed in Table~\ref{th2olvg}.

We investigated the excitation conditions of the \hho\ line emission using a radiative transfer code in the
LVG approximation \citep{ceccarelli2003} and adopting a plane parallel geometry. The molecular data were taken from the BASECOL\footnote{http://basecol.obspm.fr} database \citep{dubernet2006,dubernet2013} and we used the new collisional rate coefficients with \hh\
\citep{dubernet2009,daniel2010,daniel2011}. The linewidth (FWHM of the line profile) was set to a fixed value of 10~\kms.
The model includes the effects of the beam filling factor, and it computes the reduced chi-square $\chi^2_{\rm r}$ for each column density minimizing with respect to the source size, kinetic temperature, and density.
We adopted an uncertainty in the integrated intensities of 30\,\% for all line except the HIFI lines with high $E_{\rm u}$ and the PACS lines lying at the edges of the band, for which the adopted uncertainty is 50\,\%.

\begin{table*}[!t]
\caption{Observed and predicted water line fluxes for an OPR \hho= 3. Uncertainties are indicated in parenthesis. }
\begin{center}
\begin{tabular}{lcccccccc}
\hline\hline
&&&&&
\multicolumn{3}{c}{LVG predictions}\\
\cline{6-8}\\
Line 	&Freq. 	&$\lambda$	&$E_{\mathrm{u}}/k_{B}$     	&Flux[12.7$''$]$^{\mathrm{a}}$		&Comp.\,1-Warm$^{\mathrm{b}}$	&Comp.\,2-Hot$^{\mathrm{c}}$	&Total \\
	&(GHz)     	&(\mum)		&(K)   					&(K~\kms)			 				&(K~\kms)					&(K~\kms)		&(K~\kms)	 \\
\hline
\textbf{o-\hho} \\
\hline
$1_{10}-1_{01}$   			&\phn556.936 	&538.66		&\phn26.7     &35.0(13.0)$^{\rm{d}}$ 	&39.0\phn					&8.3	&47.3\\
$3_{12}-3_{03}$ 			&1097.365    	&273.38		&215.1	     &\phn6.0(\phn2.0)\phe\		&\phn0.3\phn					&1.0	&\phn1.3\\ 	
$3_{12}-2_{21}$ 			&1153.127    	&260.17		&215.1           &\phn1.9(\phn0.9)\phe\	&\phn0.05				          &1.4	&\phn1.5\\
$3_{21}-3_{12}$ 			&1162.912    	&257.98		&271.0	     &\phn0.8(\phn0.4)\phe\		&\phn0.02				&0.9	&\phn0.9\\
$2_{21}-2_{12} $			&1661.008    	&180.49		&159.8 	     &\phn3.0(\phn1.0)\phe\		&\phn0.2\phn					&1.6	&\phn1.8\\
$2_{12}-1_{01}	$		  	&1669.905    	&179.52		&\phn80.1	     &14.0(\phn4.0)\phe\		&\phn9.2\phn					&2.7	&11.9 \\
$3_{03}-2_{12}$			&1716.770	&174.62		&162.5	     &\phn6.0(\phn2.0)\phe\		&\phn1.0\phn					&3.4	&\phn4.4\\
$4_{14}-3_{03}$			&2640.474	&113.54		&289.3	     &\phn0.6(\phn0.3)\phe\		&\phn0.06  				&0.4	&\phn0.5\\
$2_{12}-1_{10}$			&2773.977	&108.10		&159.9	     &\phn0.6(\phn0.2)\phe\		&\phn0.4\phn					&0.6	&\phn1.0\\
\hline
 \textbf{p-\hho} 	\\
\hline
$2_{11}-2_{02}$   			&\phn752.033 	&398.92		&136.9           &11.4(\phn3.3)\phe\			&\phn1.3\phn				&6.2	&\phn7.5\\
$2_{02}-1_{11}$ 			&\phn987.927 	&303.67		&100.8           &19.0(\phn6.0)\phe\			&\phn2.9\phn				&8.4	&11.3\\
$1_{11}-0_{00}$        		&1113.343    	&269.47		&\phn53.4     &21.0(\phn6.0)\phe\			&12.8\phn					&6.3	&19.1\\
$3_{13}-2_{02}$			&2164.132	&138.54		&207.7	     &\phn1.1(\phn0.3)\phe\			&\phn0.1\phn				&1.0	&\phn1.1\\
\hline
\end{tabular}
\tablefoot{
\tablefoottext{a}{Corrected fluxes obtained for a common angular resolution of $12\farcs7$.}
\tablefoottext{b}{LVG predicted fluxes for the best-fit model of the warm component (\Tkin=250~K, $n$(\hh)=10$^{6}$~\cmt, $N$(o-\hho)=2$\times10^{14}$~\cmd, $N$(p-\hho)=$7.0\times10^{13}$~\cmd, and a source size of $10''$) and using an OPR \hh\ of 0.5 (see Sect.~4.2.1. and 4.2.2).}
\tablefoottext{c}{LVG predicted fluxes for the best-fit model of the hot component (\Tkin=1000~K, $n$(\hh)=$2\times10^{4}$~\cmt, $N$(o-\hho)=7$\times10^{16}$~\cmd, $N$(p-\hho)=$2.1\times10^{16}$~\cmd, and a source size of $2\farcs5$) and using an OPR \hh\ of 3 (see Sect.~4.2.1. and 4.2.2).}
\tablefoottext{d}{Flux after removing the contribution of the extended component detected in the 556.9~GHz line (see Sect.~3.3).}
}
\end{center}
\label{th2olvg}
\end{table*}

\subsection{Shock emission: best-fit model}

\subsubsection{A two-temperature model}

The present LVG calculations were carried out for an ortho-to-para ratio (OPR) of 0.5 in the \hh\ gas,
which is close to the value estimated by \citet{nisini2010b} for gas in the same range of excitation conditions from modelling the \hh\ emission, and an OPR of the \hho\ gas equal to 3. We also adopted a size of 10$''$, as estimated from the PACS maps, consistent with our previous findings \citep{benedettini2012} and with the {\em Spitzer} image of L1157-B1 observed in the \hh\ lines \citep{nisini2010b} and in the IRAC bands \citep{takami2011}.

\begin{table*}[!ht]
\caption{Physical conditions of the shock components accounting for the water line emission in L1157-B1.}
\begin{center}
\begin{tabular}{lccccccccc}
\hline\hline
 Comp.  & \Tkin\ 	& $n$(\hh)    & $N$(\hho)   &$N$(\hh)	& $X$(\hho) 	&  Size   &$L$(\hho) & $L$(CO) 	 &[\hho]/[CO]  \\
        &   (K)         		& (\cmt)        & (\cmd)    	&(\cmd)			 &   	&  ($\arcsec$)  &(\lo)&  (\lo)& \\ \hline
Warm    	& 250--\phn300    	& (1--3)$\times 10^6$     	& (1.2--2.7)$\times 10^{14}$  	&$1.2\times10^{20}$	& (0.7--2.0)$\times 10^{-6}$  	& 10     &0.002	& 0.004  &  0.03   \\
Hot		&900--1400    	& (0.8--2)$\times 10^4$ 		& (4.0--9.1)$\times10^{16}$   	&$3.3\times10^{20}$	& (1.2--3.6)$\times10^{-4}$	 &  2-5	&0.03\phn	& 0.01\phn & 1 \\
\hline
\end{tabular}
\end{center}
\label{h2oprop}
\end{table*}

\begin{figure}[!t]
\begin{center}
\begin{tabular}{c}
           \epsfig{file=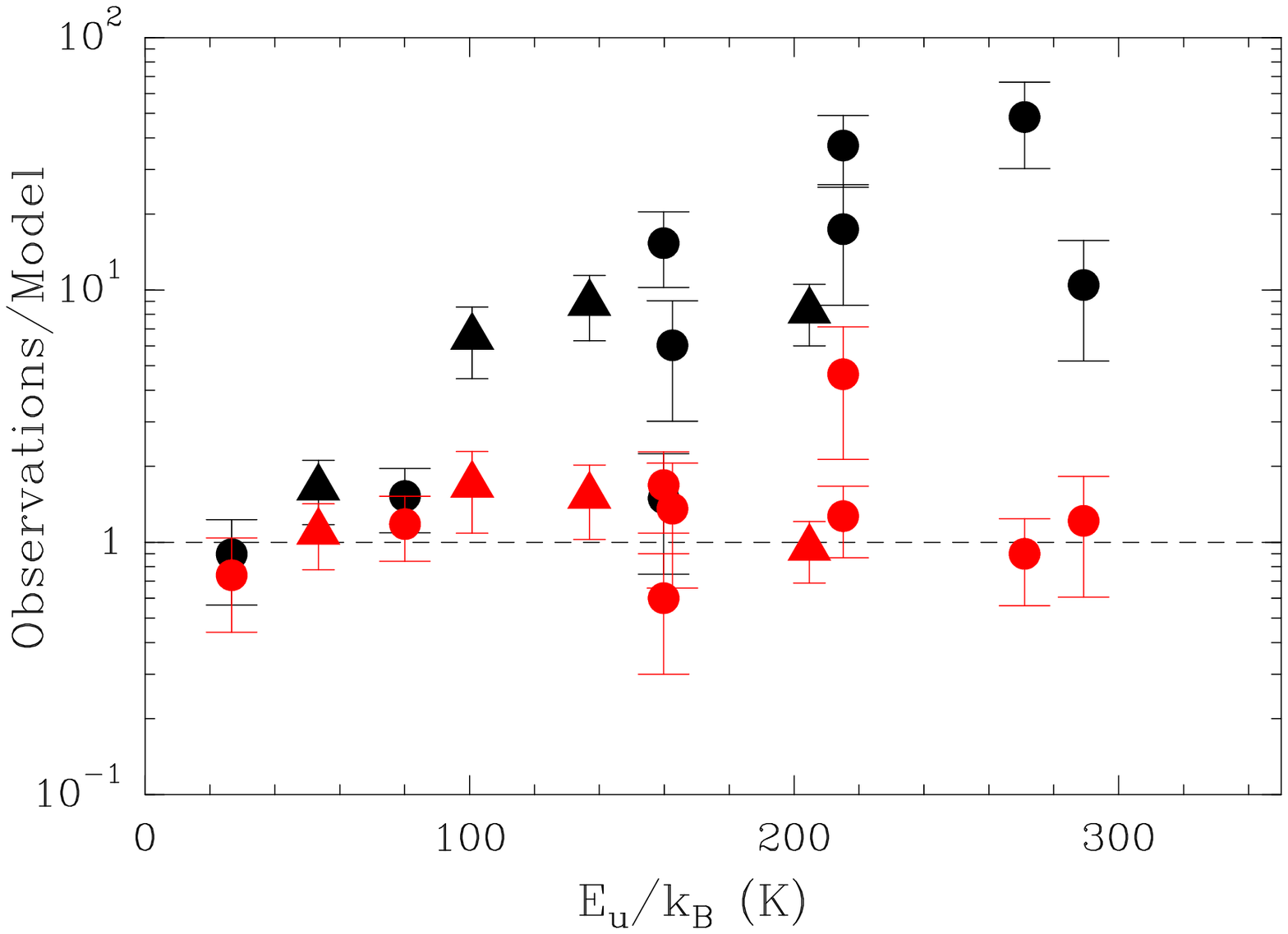,scale=0.48,angle=-90} \\
          \epsfig{file=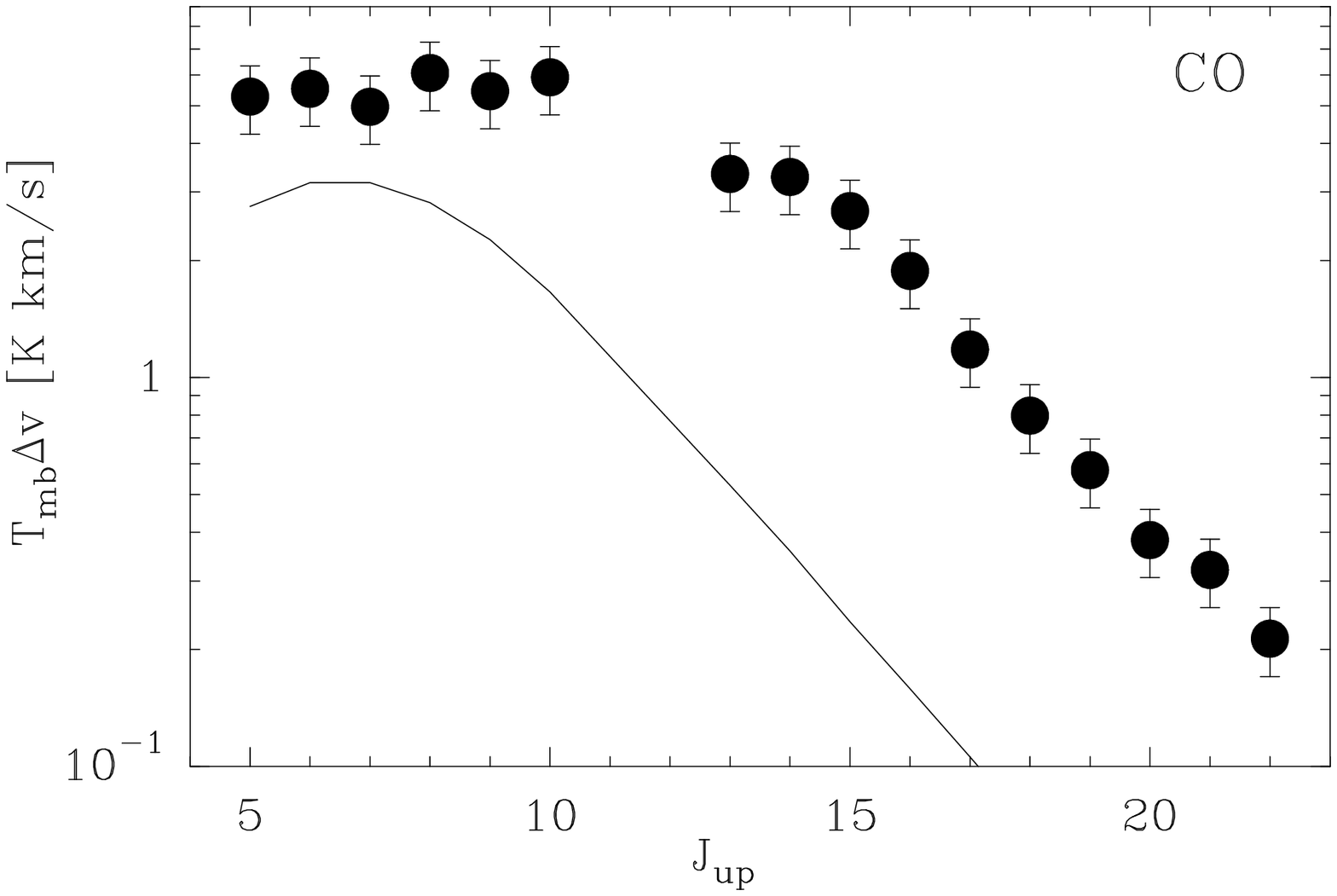,scale=0.47,angle=-90}\\
	   \end{tabular}
    \caption{{\em Top:}~ratio between the measured integrated intensities and the LVG model predictions. Filled circles/triangles depict o-\hho\ and p-\hho\ lines, respectively. In black, we display the results assuming one single temperature component (\Tkin=250~K, $n$(\hh)$=10^6$~\cmt, $N$(o-\hho)$=2\times10^{14}$~\cmd, size=10$''$). In red, the final solution when adding the contribution of the second temperature (\Tkin=1000~K, $n$(\hh)$=2\times10^4$~\cmt, $N$(o-\hho)$=7\times10^{16}$~\cmd, size=$2\farcs5$).
 {\em Bottom:}~predicted CO fluxes from the \textbf{hot} component, as a function of the rotational upper level in the HIFI and PACS range.
 The observed fluxes \citep{benedettini2012,lefloch2012} are marked with filled circles. The choice of [CO]/[\hho] =1 provides an upper limit estimate of the CO integrated intensity (see Sect.~4.2.3 for more details).}
\label{lvgbf}
\end{center}
\end{figure}

To account for the line fluxes of the three transitions connected to the ground state (ortho and para) and assuming a source size of 10$''$, the acceptable range of physical condition is \Tkin$\sim$200-300~K, $n$(\hh)=$(1-6)\times10^6$~\cmt, and $N$(o-\hho)=$(0.8-2)\times10^{14}$. The best-fit model yields
a warm gas component at 250~K, $n$(\hh)=$1\times10^6$~\cmt, $N$(o-\hho)=$2\times10^{14}$~\cmd. These physical conditions are absolutely unable to account for the flux of lines at higher upper energy levels (see Fig.~\ref{lvgbf}-top panel). 

A second gas component, at a much higher temperature and lower density, is needed to reproduce the flux detected in  the higher $E_{\rm u}$ transitions. Solutions with \Tkin$\simeq$650~K,
$N$(o-\hho)=$1\times10^{17}$~\cmd, $n$(\hh)=$8\times10^3$~\cmt, and a source size of $4\arcsec$, are possible, in principle,  under the assumption that the OPR-\hh\ remains unchanged, equal to 0.5.
Observational constraints on \hh\ suggest a higher value of OPR-\hh, typically $\simeq3$, at high temperatures (see below).
Assuming a typical OPR-\hh\ of 3 for the second, hot gas component, our modelling favors higher-temperature solutions, with
\Tkin$\simeq$1000~K, $N$(o-\hho)=$7\times10^{16}$~\cmd, $n$(\hh)=$2\times10^4$~\cmt, and a size of $2\farcs5$ as a best-fit model. We found, however, a range of possible solutions, with \Tkin$\sim$900-1400~K, $N$(o-\hho)=$(3-7)\times10^{16}$~\cmd, $n$(\hh)=$(0.8-2)\times10^4$~\cmt, and a size of 2$''$-5$''$.

{\em Therefore, a small region of hot, low density gas is contributing in addition to the warm dense $g_{\rm 1}$ 
gas, to the water emission detected by Herschel}.

To evaluate the quality of our best-fit model, we have computed the ratio of the measured water line fluxes to those predicted by our model as a function of the upper energy level of the transition. As can be seen in Fig.~\ref{lvgbf}-top panel, the overall agreement between the measured fluxes and the observations is satisfying; indeed a $\chi^2_{\rm r}$ minimization of our two-temperature model yields $\chi_{\rm r}^2=1.0$. The water line fluxes resulting from the LVG modelling are reported in Table~\ref{th2olvg} and the range of physical conditions of the warm ($g_{\rm 1}$) 
and hot shock gas components are summarized in Table~\ref{h2oprop}. The large number of lines detected at high SNR together with the availability of a wealth of complementary data allow us to constrain the water excitations conditions with unprecedented precision. It is important, however, to remark that source sizes have been imposed and this two-component model is a simplification of the complex structure of the bow-shock, in which, most likely, a wide and continuous range of temperatures and densities are present. 

Finally, we evaluated the influence of the OPR-\hho\ on the results. Only models with an OPR-\hho\ of 3 (the statistical equilibrium value) can match the observed line fluxes while a value of 1 always yields solutions with a $\chi_{\rm r}^2$\,>2.

\subsubsection{Influence of the ortho-to-para \hh\ ratio}

In their study of the emission of the pure rotational lines of \hh\ with {\em Spitzer}, \citet{nisini2010b} found evidence for two  gas components  at $\approx$300~K and 1400~K, respectively. They modeled the OPR-\hh\ as varying continuously from a value of $\approx 0.6$ in gas at 300~K to its value at LTE ($=3$) in gas at 1400~K. Therefore, we explored the range of acceptable solutions $(n, N,T)$ when considering OPR-\hh\ as a free parameter. The best-fit solution was obtained for an OPR of 0.5 in the gas at 250~K, hence a value similar to that found by \citet{nisini2010b} in the gas of moderate excitation.
We could not find any reasonable set of physical conditions 
for values of OPR-\hh\ higher than 1 for that component.

As noticed by \citet{wilgenbus2000}, such a low value of the OPR-\hh\ indicates that the gas has been recently heated up by the passage of the shock front, and not affected by an older shock episode since the timescale between shock episodes is much less than the time needed for the OPR-\hh\ to return to the equilibrium value.
This is consistent with the youth of B1, for which the estimated dynamical age is $\sim$2000~yr \citep{gueth1996} and the evolutionary age of the shock model presented by \citet{gusdorf2008}.
Low values of the OPR-\hh\ have been reported in other outflow shock regions \citep[\eg][]{neufeld1998,neufeld2006,lefloch2003,maret2009}.

As for the second, hot gas component contributing to the water line emission, higher-temperature solutions are favored when adopting an OPR-\hh\ of 3, and we found satisfying solutions ($\chi^2_{\rm r}$=0.8--1.2) for \Tkin$\simeq$1000~K,  and a gas column density $N$(\hho)$\simeq$9$\times10^{16}$~\cmd. The density and the size are less well constrained, with values of the order
a few  $10^{3-4}$~\cmt\ and a few arcsec, respectively.

\subsubsection{Modelling consistency}

Since our simple model aims at reproducing only the water line fluxes and not the line profiles, one may question its consistency with respect to the spectroscopic information of the line profiles obtained with HIFI. In other words, is there any evidence for specific observational signatures of the two temperature components invoked in our modelling?

From Fig.~\ref{lvgbf} (top panel), it appears immediately that the bulk of flux of most of the lines in the HIFI and PACS range actually comes from the hot gas component at \Tkin$\approx$1000~K. Conversely, the lines at 556.9, 1669, and 1113~GHz (HIFI) are very well accounted by the warm component at  \Tkin$\approx$250~K. This is indeed consistent with the two groups of line profiles (1113/1669~GHz and 752/998~GHz) identified (see Fig.~\ref{hificomp} in Sect.~3.2). The line profiles are very similar within each group, and differ markedly
from one group to the other. Our model provides a simple explanation to this observational fact: we are actually probing two different regions with different excitation conditions.

\begin{figure}[!t]
\begin{center}
\begin{tabular}{c}
           \epsfig{file=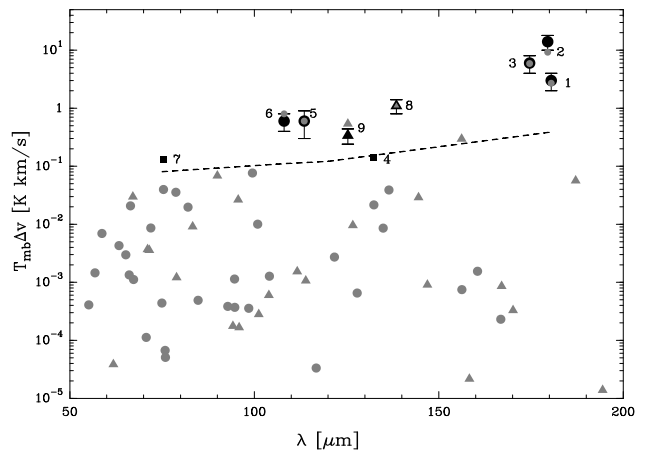,scale=0.35,angle=-90} \\
	   \end{tabular}
    \caption{Predicted \hho\ fluxes in the PACS range of the two-component model, shown in Fig.~\ref{lvgbf} and Table~\ref{th2olvg}, as a function of wavelength. Black/grey circles and triangles mark the observed/predicted fluxes of o-\hho\ and p-\hho\ lines, respectively. Squares represent upper limits of \hho\ lines listed in Table~\ref{tpacsflux}. 
    The line\_ID is also shown (see Table~\ref{tpacsflux}). The dashed line indicates the sensitivity limit of PACS.}
\label{lvg-pacsrange}
\end{center}
\end{figure}

Second, we have compared our PACS observations with the fluxes predicted by our two-temperature model for all the water lines falling in range 50--200~\mum. As can be seen in Fig.~\ref{lvg-pacsrange}, most of the lines remain below the dashed line, which draws the sensitivity limit of the observations. Our model does not predict more lines lying above the sensitivity limit than those actually detected.

CO line observations with PACS and HIFI revealed a warm, dense gas component, {\em thermalized  at 220~K}, which \citet{benedettini2012} and \citet{lefloch2012} attributed to the jet impact region against the B1 cavity. Both the location and the temperature of this component agree with the properties with the warm gas component identified by  \citet{nisini2010b}. 
However, since the bulk of emission of the CO\,(16--15) and \hho\ 1097~GHz lines arises from two components of different excitation inside the B1 cavity, we conclude that the profile of the
\hho\ 1097~GHz line could actually not be specific of $g_{\rm 1}$, unlike claimed in a previous work \citep{lefloch2012}, indicating a more complex origin of that spectral feature.

One may wonder why the CO counterpart of the second component is not detected by the sensitive PACS and HIFI instruments. This is illustrated in the bottom panel of Fig.~\ref{lvgbf}, which displays the predicted CO fluxes for the hot component (\Tkin$\approx$1000~K, $n$(\hh)=$2\times10^4$~\cmt, and the assumed size of $2\farcs5$), assuming an abundance ratio
[CO]/[\hho]=1. 
Adopting the standard value of $10^{-4}$ for CO would imply a smaller value of the column density, and the flux of the hot component
would be even smaller.

We point out that a similar two-component model has been presented recently by \citet{santangelo2013} to account for the water emission towards the B2 shock position of the L1448 molecular outflow, where the physical conditions are similar to the ones obtained in L1157-B1.
Confirming the presence of similar two-component structure in other shock regions would suggest that both components are most likely related to the bow-shock phenomenon itself. The nature of the relation could provide some clue onto the origin of the line profiles observed.

\subsection{Cloud absorption}

In Sect.~3.2, we showed that \hho\ transitions connecting with the ground state level present a narrow self-absorption feature close to the ambient velocity, at \vel$\sim$2.6~\kms.
Since optically thin lines, such as \tco\ (Lefloch \et\ in preparation) and HDO \citep{codella2012}, peak close to the ambient velocity, we propose that the absorption feature seen in the low-excited \hho\ lines most likely arises from an extended layer associated with the cloud envelope, as a result of  ice photodesorption. Such a model was  successfully applied recently by
\citet{coutens2012} to the low-mass Class\,0 protostar IRAS\,16293--2422, where the authors find a similar self-absorption signature in the fundamental lines of HDO and H$_{2}^{18}$O. In order to account for  the observed line profiles, the authors added an absorbing layer in front of the IRAS\,16293--2422 envelope that results from the photodesorption of icy mantles at the edge of the cloud by the FUV photons, as modeled by \citet{hollenbach2009}.

Interpreting in a similar way the absorption feature as due to a water-rich layer caused by ice photodesorption at  the cloud surface, we can estimate its water abundance.
\citet{caratti2006} evaluate the visual extinction towards the B1 shock through near-IR data and find that $A_{\rm V}$ is, at most, 2 mag. Thus, adopting  $A_{\rm V}$ of 1--2~mag, assuming an incident FUV flux of $G_{0}$=1 (\ie\ adopting a standard interstellar radiation field), a typical cloud gas density of 10$^4$~\cmt, \citet{hollenbach2009} predict a water abundance of about $\sim 10^{-7}$. At the same depth and for a fixed value of $G_{\rm 0}$, lower densities
would result in slightly lower values of the \hho\ abundance. On the other hand, if we consider higher values of $G_{\rm 0}$ the water abundance, for $A_{\rm V}$ of 1--2~mag, will be lower as a higher incident flux modifies the depth of the freeze, moving it towards higher visual extinction, resulting in a peak deeper in the cloud for gas-phase \hho. Therefore, we estimate the water abundance of the cloud absorbing layer due to the cloud $\simeq$10$^{-7}$.
Adopting the relation $N$(\hh)=9.4$\times10^{20}$\,$A_{\rm V}$ \citep{frerking1982} we obtained $N$(\hh)=(0.9-1.9)$\times10^{21}$~\cmd\ for an $A_{\rm V}$ of 1--2~mag, and hence the column density of \hho\ in this absorbing layer should be about 1$\times10^{13}$~\cmd.

\subsection{Water abundance and line cooling}

The water abundance $X$(\hho)=[\hho]/[\hh] in the shocked gas was derived using the \hh\ data from \citet{nisini2010b} obtained with {\em Spitzer} and  convolved at the PACS resolution ($12\farcs7$).
For the 250~K gas component, 
the \hh\ column density was computed  from an LTE analysis of the S(0) to S(2) rotational lines. For a source size of 10$''$, the column density of \hh\ is $N$(\hh)$\simeq 1.2\times 10^{20}\cmd$, which yields a fractional abundance of water $\sim$(0.7--2)$\times 10^{-6}$. This warm gas component is associated with a partly dissociative J-type shock either with a shock velocities and pre-shock densities of \velo$>$30~\kms\ and $2\times10^{4}$~\cmt\ or \velo$>$20~\kms\ and $2\times10^{5}$~\cmt\ \citep{benedettini2012,lefloch2012}, and hence the low water abundance can be explained in terms of FUV photons produced at the shock front that prevent the full conversion of free oxygen into water, resulting in a decrease of the water abundance. To obtain the water abundance of the hot gas, we considered the \hh\ rotational lines S(5) up to S(7) and computed the \hh\  column density scaled for a source size of size of $2\farcs5$. The derived water abundance is (1.2--3.6)$\times10^{-4}$, in agreement with the predicted values for hot shocked material \citep{kaufman1996,bergin1998,flower2010}.
The abundance of \hho\ in the hot gas is two orders of magnitude higher than that obtained for the warm gas, indicating that all of the available oxygen not locked in CO has been converted to \hho. The derived \hho\ abundances are reported in Table~\ref{h2oprop}.

The low water abundance associated with the warm gas component confirms previous findings in molecular outflows based on a limited number of lines \citep[\eg][]{bjerkeli2012,vasta2012,tafalla2013,santangelo2013}. Moreover, we also confirmed the presence of a hot gas component at higher abundance that so far has been clearly identified only for the B2 shock position of the L1448 outflow \citep{santangelo2013}. The presence of a warm and hot water components have been suggested by \citet{goicoechea2012} and \citet{dionatos2013} in shocks close to several Class~0 sources in Serpens. Therefore, our results confirm that in bow-shocks far from the driving source there is a bimodal distribution, which seems to be a common shock characteristic.

\citet{nisini2010b} obtained the line cooling due to \hh\ in the B1 shock position, which is of the order of 0.03~\lo\footnote{$L_{\rm H_{2}}$ has been corrected for a distance of 250~pc while in \citet{nisini2010b} the adopted distance is 440~pc.}.
Here, we estimated the total luminosity of water lines, $L_{\rm H_{2}O}$, from the predicted line fluxes of the best-fit model shown in Fig.~\ref{lvgbf}.
We obtained $\sim$0.002~\lo\ and $\sim$0.03~\lo\ for the warm and the hot gas components, respectively. Regarding CO, the derived luminosity of the warm component is 0.004~\lo, hence the contribution of water to the line cooling is 50\,\% of the CO luminosity. The luminosity of hot CO gas component, on the other hand, is 0.01~\lo, and therefore the far-IR cooling of \hho\ dominates in front of CO, and it contributes equally as the \hh\ line cooling. The results are summarized in Table~\ref{h2oprop}.

Finally, we calculated the total far-IR cooling in the B1 shock region following the definition of \citet{nisini2002}, where $L_{\rm{FIR}}=L_{\rm{OI}}+L_{\rm{CO}}+L_{\rm{H_{2}O}}+L_{\rm{OH}}$. Using the line fluxes reported by \citet{benedettini2012}, we estimated the total luminosity of [OI] and OH , which are $L_{\rm{OI}}$\,$\simeq$2$\times10^{-3}$~\lo\ and $L_{\rm{OH}}$\,$\simeq$4$\times10^{-4}$~\lo. For CO and \hho\ we considered the contribution of the two gas components.
One can clearly see that the far-IR cooling is dominated by the contribution of \hho\ and CO lines, followed by [OI] and OH. The total far-IR cooling estimated in B1 is $\sim$0.05~\lo. It is worth noticing that the FIR luminosity has been computed using similar beam sizes for all species, while the \hh\ luminosity was estimated with a smaller beam.

Shock models produce markedly different predictions on the \hho\ cooling function, depending on the nature of the shock, either C-type (MHD) or J-type shocks. We make here a simple comparison with the predictions from the steady-state shock models of \citet{flower2010}, for a shock propagating at \velo=20~\kms\ into gas with pre-shock density of $10^4$~\cmt. Our goal is to identify qualitative trends on the properties of the shock responsible for the \hho\ emission detected. As pointed out by \citet{gueth1996}, the B1 bow-shock is propagating into gas previously accelerated by the ejection associated with B2. Maximum velocities of $5-10$~\kms\ are reported in the B2 outflow cavity \citep{vasta2012}. For this reason, we consider that velocities detected in the \hho\ gas towards B1 ($\approx$30~\kms; Fig.~\ref{waterprofiles}) are not inconsistent with a shock velocity of about 20~\kms.

For  the warm (\Tkin$\simeq$250~K) 
gas component, the \hho\ line cooling is $\approx$1.5$\times10^{-19}$~erg\,\cmt\,${\rm s^{-1}}$, in agreement with the value predicted in the molecular reformation zone of a J-type shock (see Fig.~2 of \citealt{flower2010}; see also \citealt{benedettini2012}).
For the hot (\Tkin$\simeq$1000~K) gas component, the \hho\ line cooling is $\approx$2$\times10^{-16}$~erg\,\cmt\,${\rm s^{-1}}$, a value several order of magnitude higher than that predicted by the C-shock model, but well in the range of values expected in the J-type shock.
Therefore, the simple comparison suggests that the hot gas layer is excited in a non-dissociative J-type shock.

\section{Summary and Conclusions}

As part of the CHESS key program, we have analyzed the \hho\ emission towards the shock region L1157-B1. A grand total of 13 \hho\ lines (both ortho and para) have been detected with HIFI and PACS instruments arising from transitions with rather low $E_{\rm u}$, from 26.7~K to 319.5~K.
The PACS and HIFI observations towards the L1157-B1 bow-shock have revealed the presence of several gas components with different excitation. Our main conclusions can be summarized as follows~:

\begin{enumerate}

\item The bulk of \hho\ emission originates in the B1 bow-shock.

\item An absorption feature is detected in the line profiles connecting with the ground state level. It arises from a water-enriched layer ($X$[\hho]$\simeq$10$^{-7}$) at the surface of the cloud formed as a result of water ice photodesorption from interstellar grain mantles,  driven by the external UV photons due to the interstellar radiation field.

\item The LVG analysis of the \hho\ emission associated with the bright high-excitation region (\ie\ the bow-shock) has permitted us to identify two physical components. 
A warm, dense gas (\Tkin$\sim$200-300~K, $n$(\hh)$\simeq$(1--3)$\times10^6$~\cmt) component traced mainly by the low-excitation lines of water (shown in blue in Fig.~\ref{cartoon}), with an assumed extent of 10$''$. The OPR-\hh\ in the warm gas is  $\simeq$0.5. The hot (\Tkin$\simeq$1000~K) component is made of tenuous gas at a much lower density (a few 10$^{3-4}$~\cmt) similar to that of the parental cloud. It is much more compact,  with a typical size of  $2''-5''$. The OPR-\hh\ in the warm gas is  $\simeq$3.0, equal to its value at LTE.  

\item These two shock components present marked  differences in terms of water enrichment. While the derived abundance in the warm gas is (0.7--2)$\times10^{-6}$, the water abundance estimated in the hot gas is much higher, around (1.2--3.6)$\times10^{-4}$, indicating that all available oxygen not locked in CO is driven into \hho. The FIR cooling of the bow-shock appears to be equally dominated by both \hh\ and the hot water component.

\item A simple comparison of the water line cooling properties with the steady-state shock models of \citet{flower2010} is consistent with a J-type shock origin for both components. The exact nature of the hot water spot and its relation with the jet impact against the cavity remains to be established. 
The low density of the hot \hho\ gas suggests that the shock propagates into a region of much lower density, either in the ambient cloud or the outflow cavity gas.

\end{enumerate}

Higher-angular observations are needed to understand the structure of the L1157-B1 bow-shock region. We expect that a detailed, multiline study and comparison of the emission properties of the major cooling agents, CO, \hho\ and \hh\ at infrared wavelengths with shock model predictions, will help us to clarify the origin and the relation the different shock components revealed by PACS and HIFI hold to each other  (Cabrit et al. in preparation).

\begin{acknowledgements}

The authors are grateful to the anonymous referee and the Editor Dr. Malcolm Walmsley for valuable comments. G.B is grateful to M. Pereira-Santaella for fruitful discussion.
G. Busquet, M. Benedettini, C. Codella, B. Nisini, A.~I. G\'omez-Ruiz, and A.~M. di Giorgio are supported by the Italian Space Agency (ASI) project I/005/11/0.
B. Lefloch thanks the Spanish MEC for funding support through grant SAB2009-0011.
B. Lefloch, C. Ceccarelli, and L.Wiesenfeld acknowledge funding from the French Space Agency CNES and
the National Research Agency funded project FORCOM, ANR-08-BLAN-0225. S. Viti acknowledges support from the [European Community's] Seventh Framework Programme [FP7/2007-2013] under grant agreement n$\degr$ 238258. A. Gusdorf acknowledges support by the grant ANR-09-BLAN-0231-01 from the French Agence Nationale de la Recherche as part of the SCHISM project.
HIFI has been designed and built by a consortium of institutes and university departments from across Europe, Canada and the United States under the leadership of SRON Netherlands Institute for Space Research, Groningen, The Netherlands and with major contributions from Germany, France and the US. Consortium members are: Canada: CSA, U.Waterloo; France: CESR, LAB, LERMA, IRAM; Germany: KOSMA, MPIfR, MPS; Ireland, NUI Maynooth; Italy: ASI, IFSI-INAF, Osservatorio Astrofisico di Arcetri-INAF; Netherlands: SRON, TUD; Poland: CAMK, CBK; Spain: Observatorio Astron\'omico Nacional (IGN), Centro de Astrobiolog\'ia (CSIC-INTA). Sweden: Chalmers University of Technology - MC2, RSS \& GARD; Onsala Space Observatory; Swedish National Space Board, Stockholm University - Stockholm Observatory; Switzerland: ETH Zurich, FHNW; USA: Caltech, JPL, NHSC. PACS has been developed by a consortium of institutes led by MPE (Germany) and including UVIE (Austria); KU Leuven, CSL, IMEC (Belgium); CEA,!
  LAM (France); MPIA (Germany); INAF-IFSI/OAA/OAP/OAT, LENS, SISSA (Italy); IAC (Spain). This development has been supported by the funding agencies BMVIT (Austria), ESA-PRODEX (Belgium), CEA/CNES (France), DLR (Germany), ASI/INAF (Italy), and CICYT/MCYT (Spain).

\end{acknowledgements}

\bibliography{l1157-h2o-accepted}

\end{document}